\documentclass[11pt,twoside,letterpaper]{article} 
\usepackage{times,fancyhdr}
\usepackage[dvips]{graphicx}

\usepackage{epsfig}
\usepackage{amssymb}
\usepackage{amsmath}
\usepackage{amsfonts}
\usepackage{amsthm,amscd}
\usepackage{amsbsy}
\usepackage{latexsym}
\usepackage{bm}
\usepackage{url} 			
\usepackage{layout}
\usepackage{pslatex}
\usepackage{cite}
\usepackage{fleqn} 		
\usepackage{makeidx}
\makeindex 						
\usepackage{layout} 	
\usepackage{epstopdf}	
\usepackage{color}
\usepackage{hyperref}
\usepackage[latin1]{inputenc}
\usepackage[T1]{fontenc}
\usepackage{poligraf} 
\usepackage[letter,cam,center]{crop} 
\usepackage{type1cm} 	
\usepackage{courier}	
\usepackage{lscape} 	


\makeatletter 
\def\cleardoublepage{\clearpage\if@twoside \ifodd\c@page\else%
    \hbox{}%
    \thispagestyle{empty}%
    \newpage%
    \if@twocolumn\hbox{}\newpage\fi\fi\fi} 
\makeatother

\def\figurename{Figure}
\makeatletter
\renewcommand{\fnum@figure}[1]{\figurename~\thefigure.}
\makeatother

\def\tablename{Table}
\makeatletter
\renewcommand{\fnum@table}[1]{\tablename~\thetable.}
\makeatother

\begin{document}
\title{
{\begin{flushleft}
\vskip 0.45in
{\normalsize\bfseries\textit{Chapter~1}}
\end{flushleft}
\vskip 0.45in
\bfseries\scshape Spin dynamics in semiconductors in the streaming regime}}
\author{\bfseries\itshape L.~E.~Golub\thanks{E-mail address: golub@coherent.ioffe.ru} \: and \: E.~L.~Ivchenko\\
Ioffe Physical-Technical Institute of the Russian Academy of Sciences, 194021
St.~Petersburg, Russia}
\date{}
\maketitle
\thispagestyle{empty}
\setcounter{page}{1}
\thispagestyle{fancy}
\fancyhead{}
\fancyhead[L]{In: Book Title \\ 
Editor: Editor Name, pp. {\thepage-\pageref{lastpage-01}}} 
\fancyhead[R]{ISBN 0000000000  \\
\copyright~2014 Nova Science Publishers, Inc.}
\fancyfoot{}
\renewcommand{\headrulewidth}{0pt}

\vspace{2in}

\noindent \textbf{PACS}
72.25.Hg,	
72.25.Pn,	
72.25.Rb,	
73.63.Hs. 

\vspace{.08in} \noindent \textbf{Keywords:} Spin-orbit splitting, semiconductor heterostructures, streaming

\begin{abstract}
We present results of a cross-disciplinary theoretical research at the interface of spin physics and hot-electron transport. A moderately strong electric field is assumed to provide the streaming regime
where each free charge carrier, an electron or a hole, accelerates quasiballistically in the ``passive'' region until reaching the optical-phonon energy, then emits an optical phonon and starts the next period of acceleration. The inclusion of spin degree of freedom into the streaming-regime kinetics gives rise to rich and interesting spin-related phenomena. Firstly, in the streaming regime the spin relaxation is substantially modified, and the current-induced spin orientation is remarkably increased. Under short-pulsed photoexcitation at the bottom of conduction band the photoelectrons 
execute a periodic damped motion in the energy space with the period equal to the free flight time of an electron in the passive region. If the short optical pulse is circularly polarized so that the photocarriers are spin oriented, then the spin energy distribution is oscillating in time as well, which can be detected in the pump-probe time-resolved experiments. We show that the spin-orbit splitting of the conduction band becomes a source for additional spin oscillations, periodic or aperiodic depending on the value of electric field.
\end{abstract}

\section{Introduction}

Spin physics is a rapidly growing area of research in condensed matter science aimed at the creation, manipulation and detection of spins in various systems. Important and interesting fundamental results, also promising for possible future applications, have been obtained in semiconductors and semiconductor nanostructures~\cite{Dyakonov_book}. Highly sensitive methods of Kerr and Faraday rotation spectroscopy give a possibility to probe spin orientation within subpicosecond resolution~\cite{KerrFaraday}. A challenging problem in the spin physics is how to affect the spin by instantaneous non-magnetic methods, in particular, by static or optical electric fields. In semiconductors, optical selection rules allow one to create spin-oriented carriers (optical orientation) which may then be accelerated by a static external electric field. Another possibility to monitor the spin by electrical means arises due to the spin-orbit interaction which couples the electron orbital and spin degrees of freedom. In semiconductor heterostructures, the spin-orbit interaction is described by the Rashba and Dresselhaus contributions to the electron effective Hamiltonian linear in the electron quasimomentum $\bm p$, for review see Ref.~\cite{BIA_SIA_review}. Generally, they can be presented as follows
\begin{equation} \label{H_SO}
H_{\rm so}(\bm p) = \beta_{ij} \sigma_i p_j= 	{\hbar \over 2} \bm \sigma \cdot \bm \Omega_{\bm p}\:,
\end{equation}
where $\sigma_i$ ($i=x,y$) are the Pauli matrices, $\bm \beta$ is a second-rank pseudotensor, and the effective momentum-dependent Larmor frequency is defined by $\Omega_{{\bm p}, i}=2 \beta_{ij} p_{j} /\hbar$. Hereafter, for definiteness, we consider zinc-blende-lattice based quantum wells grown along the direction [001] and use the coordinate frame $x \parallel [1\bar{1}0], y \parallel [110]$ and $z \parallel [001]$.

We study here the electron spin dynamics in semiconductor heterostructures in the presence of an external in-plane electric field. The attention is focused on the so-called streaming regime~\cite{Andronov_review,streaming_QWR,Tulupenko,Kim_streaming,Korotyeyev_streaming,Kochelap_streaming} characterized by the periodic optical-phonon emission by electrons subjected to acceleration in the strong electric field. A number of interesting time-resolved spin effects are shown to be caused by the electric field. We also discuss a possibility to detect them in the pump-probe experiments, both in the absence and presence of an external magnetic field.

\section{Streaming in 2D semiconductors}

In an external electric field $\bm F$, the carrier distribution is anisotropic in the momentum space. In particular, it is characterized by a nonzero electron drift momentum 
$p_{\rm dr}$ in the opposite to field direction.
For comparatively weak fields $F \ll 1$~kV/cm, the anisotropy is weak: the
ratio $p_{\rm dr} / p$ is smaller than $0.1$, where $p$ is the root-mean-square of the electron quasimomentum in the absence of the field.
 An application of the increasing field leads to a transformation from the linear dependence $p_{\rm dr} \propto F$ to the saturated behaviour.
If the interparticle collisions play a minor role in the kinetics, the electron momentum-space distribution in strong electric fields becomes extremely anisotropic, \textit{streaming}-like. Each electron accelerates quasiballistically in the ``passive'' region until reaching the threshold energy equal the optical-phonon energy $\hbar \omega_0$ and the corresponding threshold quasimomentum $p_0 = \sqrt{2 m \hbar \omega_0}$, where $m$ is the electron effective mass. Then it looses its energy by emitting an optical phonon and starts the next period of acceleration.

Thus, in the streaming regime realized in an appropriate range of the dc electric fields, the electron distribution in the momentum space has a steaming-like, or needle-like, form spread between the $\Gamma$-point ${\bm p}=0$ and the point ${\bm p}_0 = p_0 \hat{\bm e}$, where $\hat{\bm e}$ is the unit vector in the direction opposite to the electric field $\bm{F}$. The formation of such an anisotropic electron distribution requires the following hierarchy of times
\begin{equation} \label{tauee}
\tau \ll t_{\rm tr} \ll \tau_p, \, \tau_{\rm ee}\:.
\end{equation}
Here $\tau$ is the time of optical phonon emission by an electron in the ``active'' energy region exceeding $\hbar \omega_0$, $t_{\rm tr} = p_0/|e F|$ is the travel time through the passive region of the momentum space, $\tau_p$ is the momentum relaxation time due to electron scattering by acoustic phonons or static imperfections, and $\tau_{\rm ee}$ is the electron-electron collision time. The electron-electron scattering tends to convert the anisotropic distribution to the shifted
Maxwellian distribution with an effective electron temperature and a drift velocity $p_{\rm dr}/m$. The corresponding time can be estimated by 
$\tau^{-1}_{\rm ee} \sim (e^2/\varkappa\hbar)^2 N/\omega_0$, where $N$ is the electron 2D density and $\varkappa$ is the dielectric constant.

The electron distribution function $f_{\bm p}$ is given by~\cite{NJP}
\begin{equation}
\label{spin_independ}
f_{\bm p} = N{2\pi^2 \hbar^2\over p_0} \tilde{\delta}(p_y) 
\left\{
\begin{array}{cc}
\tilde{\theta} \left( p_x \right), &{\rm if}\hspace{3 mm} p<p_0, \\
{\rm exp} \left( - \alpha {p_x - p_{0x} \over p_0}\right), &{\rm if}\hspace{3 mm}p>p_0,
\end{array}
\right. 
\end{equation}
where $\alpha = p_0/|eF|\tau$, $p_{0x} = \sqrt{p_0^2 - p_y^2}$, and 
\[ \tilde{\delta}(p_y) = \sqrt{\alpha \over 2\pi} \frac{1}{p_0}
{\rm exp}\left( - \frac{\alpha}{2} \frac{p_y^2}{p_0^2}\right) \:,
\qquad
\tilde{\theta} \left( p_x \right) = \frac12\left[1 + {\rm erf} \left(\sqrt{\frac{\alpha}{2}} \frac{p_x}{p_0} \right) \right]\:.
\]
One can see from Eq.~(\ref{spin_independ}) that the penetration $\delta p_x$ into the active region and the width $2 \delta p_y$ of the distribution in the transverse direction $p_y$ are given by $\delta p_x = p_0/\alpha$ and $\delta p_y = p_0/\sqrt{\alpha}$. Thus, the needle-like distribution is formed provided $\alpha \gg 1$ which establishes the upper limit for the electric field strength. In this case the function $\tilde{\theta} \left( p_x \right)$ can be approximated by the Heaviside  function $\theta(p_x)$ and $p_{0x}$ by $p_0 - (p_y^2/2 p_0)$. It is also worth to mention that the value $\alpha$ gives the ratio between the numbers of particles in the passive and active regions as well as the ratio $t_{\rm tr}/\tau$ of the times spend by electrons in these regions.

If the electrons are spin polarized with the average spin ${\bm s}$ per particle, their spin distribution function in the ${\bm p}$-space is given by the main contribution
\begin{equation}
\label{S_p_f}
	\bm S_{\bm p} = 2 f_{\bm p} \bm s
\end{equation}
and an additional term of the order of $\Omega_{\bm p} t_{\rm tr}$.

\subsection{Effect of elastic scattering on streaming distribution}

In order to analyze consequences of elastic scattering by static imperfections, impurities and interface roughness, and/or quasi-elastic scattering by acoustic phonons, we start with the approximation of ``ideal streaming'' replacing the smoothed delta-function $\tilde{\delta}(p_y)$ in Eq.~(\ref{spin_independ}) by the Dirac delta-function and the smoothed theta-function $\tilde{\theta}$ by the Heaviside step function. Then the streaming distribution in the passive region reduces to
\begin{equation} \label{zero}
f^0_{\bm p}=N{2\pi^2 \hbar^2\over p_0} \delta(p_y)\theta(p_x)\theta(p_0-p_x).
\end{equation}
In the following we include into consideration the elastic scattering by short-range imperfections and describe the scattering process ${\bm p}' \to {\bm p}$ by the rate
$W_{\bm p \bm p'} = (\pi \tau_p)^{-1} \delta(p^2-p'^2)$ normalized on the inverse momentum relaxation time $\tau_p$ as follows
\[
\tau_p^{-1}= \int d{\bm p'} \, W_{\bm p \bm p'}\:.
\]
Since the streaming regime is achieved provided $\tau_p$ is much longer than the travel time $t_{\rm tr}$ we treat the elastic scattering keeping the first order corrections in
$t_{\rm tr}/\tau_p$ to the electron distribution function (\ref{zero}). Then the scatter of electrons over all azymuthal angles $\theta$ gives rise to a correction $f^{(1)}_{\bm p}$ which, in the passive region, satisfies the equation
\begin{equation} \label{first}
	eF{\partial f^{(1)}_{\bm p} \over \partial p_x} = \int d{\bm p'} \, W_{\bm p \bm p'} f^0_{\bm p'}
\end{equation}
with the boundary condition $f^{(1)}_{\bm p} = 0$ at $p_x = - p_{0x} \equiv -\sqrt{p_0^2 - p_y^2}$. Here the right-hand side term represents the generation of electrons to the whole area $p = \sqrt{p_x^2 + p_y^2} \leq p_0$ due to the elastic scattering from the zeroth-approximation needle (\ref{spin_independ}) while the term on the left-hand side describes the further acceleration of the scattered electrons in the electric field. 

Substituting Eq.~\eqref{zero} into Eq.~(\ref{first}) we obtain
\[
eF{\partial  f^{(1)}_{\bm p}  \over \partial p_x} = { \pi \hbar^2 N \over \tau_p p_0 p}\:.
\]
The solution of this first-order differential equation leads to the formula
\begin{equation}
\label{delta_f}
	f^{(1)}_{\bm p}   = {t_{\rm tr}\over  \tau_p}{\pi \hbar^2 N \over p_0^2} 
	\int\limits_{-p_{0x}}^{p_x} {dp_x'\over \sqrt{p_x'^2+p_y^2}} 
	= {t_{\rm tr}\over  \tau_p} {\pi \hbar^2 N \over p_0^2} \ln{\left({p+p_x\over p_0 - p_{0x}} \right)}.
\end{equation}
The steady-state density of electrons scattered off the needle in the passive area is given by
\begin{equation}
	\delta N = N {t_{\rm tr}\over  \tau_p} \int\limits_0^{2\pi} {d\theta \over 2\pi} \int\limits_0^1 ds \, s 
	\ln{\left[ s(1+\cos{\theta}) \over 1-\sqrt{1-s^2\sin^2{\theta}} \right]}
	= N {t_{\rm tr}\over  \tau_p} {2G+1\over \pi} \approx 0.9 \, N {t_{\rm tr}\over  \tau_p}\:,
\end{equation}
where $G \approx 0.9159\dots$ is the Catalan constant. These electrons reach the sphere ${p_x^2 + p_y^2 = p_0^2}$, emit optical phonons and return 
to the point $p=0$
Thus, in the first approximation, the total density $N$ should be replaced by $N - \delta N$.
The perturbation procedure can be continued by the replacements $f^0_{\bm p} \to f^0_{\bm p} + f^{(1)}_{\bm p}$ and $f^{(1)}_{\bm p} \to f^{(2)}_{\bm p}$
in the right- and left-hand sides of Eq.~(\ref{first}), respectively.

\section{Spin relaxation}

The constriction of the distribution function $f^0_{\bm p}$ along the $p_y$ axis and the electron homogeneous occupation of the $p_x$ interval between 0 and $p_0$ substantially modifies the Dyakonov-Perel spin relaxation of the $s_y$ component controlled by the linear-${\bm p}$ term $\beta_{xy} \sigma_x p_y$ in Eq.~(\ref{H_SO}). 
 Here, as above, we assume the electric field to be applied along the $x$ axis.
The spin-relaxation time $\tau_{sy}$ of this component is found from the equation, see~\cite{NJP}:
\begin{equation} \label{sytausy}
	{s_y\over \tau_{sy}} = {1\over N} \sum_{\bm p} \Omega_{\bm p, x} \delta S_{{\bm p},z}\:,
\end{equation}
where $\Omega_{{\bm p},x} = 2 \beta_{xy} p_y/ \hbar$, ${\bm s}$ is the average electron spin per particle, $\delta \bm S_{\bm p}$ is the first-order correction (in $\bm \Omega_{\bm p}$) to the spin distribution function satisfying the spin kinetic equation, see Eq.~(22) in Ref.~\cite{NJP},
\begin{equation} \label{deltaS}
	eF {\partial \over \partial p_x} \delta \bm S_{\bm p} + 2 f^0_{\bm p} \, \bm s \times \bm \Omega_{\bm p} = 0\:.
\end{equation}
Solving this equation with the streaming distribution~\eqref{spin_independ} we obtain~\cite{NJP}
\begin{equation} \label{tausy}
	{1\over \tau_{sy}} = 2 \tau \left({\beta_{xy} p_0 \over \hbar} \right)^2\:.
\end{equation}
We remind that, in the state of equilibrium, the relaxation time $\tau_{sy}$ is given by the similar equation with $\tau p_0^2$ replaced by $\tau_p 2m \bar{\varepsilon}$, where $\bar{\varepsilon}$ is the equilibrium  value of the electron average energy. 

Now we consider effect of elastic scattering on the spin relaxation and show that even the first-order correction in the ratio $t_{\rm tr}/\tau_p  \ll 1$ can modify $\tau_{sy}$. For this purpose we replace 
$f^0_{\bm p}$ in Eq.~(\ref{deltaS}) by $f^{(1)}_{\bm p}$ and  find the scattering-induced correction $\delta \bm S^{(1)}_{\bm p}$. The latter is given therefore by the integral
\begin{equation}
	\delta S_{{\bm p},z}^{(1)} = s_y \frac{2 \Omega_{\bm p, x}}{eF} \int\limits_{-p_{0x}}^{p_x} dp_x' \,  f^{(1)}_{p_x',p_y}\:,
\end{equation}
which can be solved analytically to give
\begin{equation} \label{sz1}
	\delta S_{{\bm p},z}^{(1)} = s_y {t_{\rm tr}\over  \tau_p}{N \pi \hbar^2\over p_0^2}  {2 \Omega_{\bm p, x} \over eF} 
	\left[ p_0 - p + p_x \ln{\left({p+p_x\over p_0 - p_{0x}} \right)} \right].
\end{equation}
Substituting this correction to Eq.~(\ref{sytausy}) and summing over ${\bm p}$ we arrive at the contribution of the elastically scattered electrons to spin relaxation of the electron gas
\begin{equation} \label{elasttaus}
	{1\over \tau_{sy}^{\rm elast.}} = {7\over 15}{t_{\rm tr}^2 \over  \tau_p} \left({\beta_{xy} p_0 \over \hbar} \right)^2\:.
\end{equation}
In contrast to the electrons in the zeroth-approximation needle distribution~(\ref{spin_independ}) compressed along the $p_y$ axis, the scattered electrons are spread over the whole passive area. The increased precession frequency $\Omega_{{\bm p},x} = 2 \beta_{xy} p_y/\hbar$ can compensate smallness of the ratio $t_{\rm tr}/\tau_p$. For comparison of the contributions~(\ref{tausy}) and~(\ref{elasttaus})  it is instructive to sum them and present the total spin relaxation time in the following convenient form 
\begin{equation} \label{totalsr}
	{1\over \tau_{sy}} = 2 \tau \left({\beta_{xy} p_0 \over \hbar} \right)^2 \left( 1 + {7\over 30} { t_{\rm tr} \over \tau_p}\alpha\right) \:.
\end{equation}
For $\alpha=33$ and $\tau_p/t_{\rm tr}=3$, we find the contribution (\ref{elasttaus}) caused by elastic scattering to exceed that given by Eq.~(\ref{tausy}) by more than twice. 

\section{Spin orientation by electric field}

\begin{figure}[h]
\begin{center}
\includegraphics[width=0.65\linewidth]{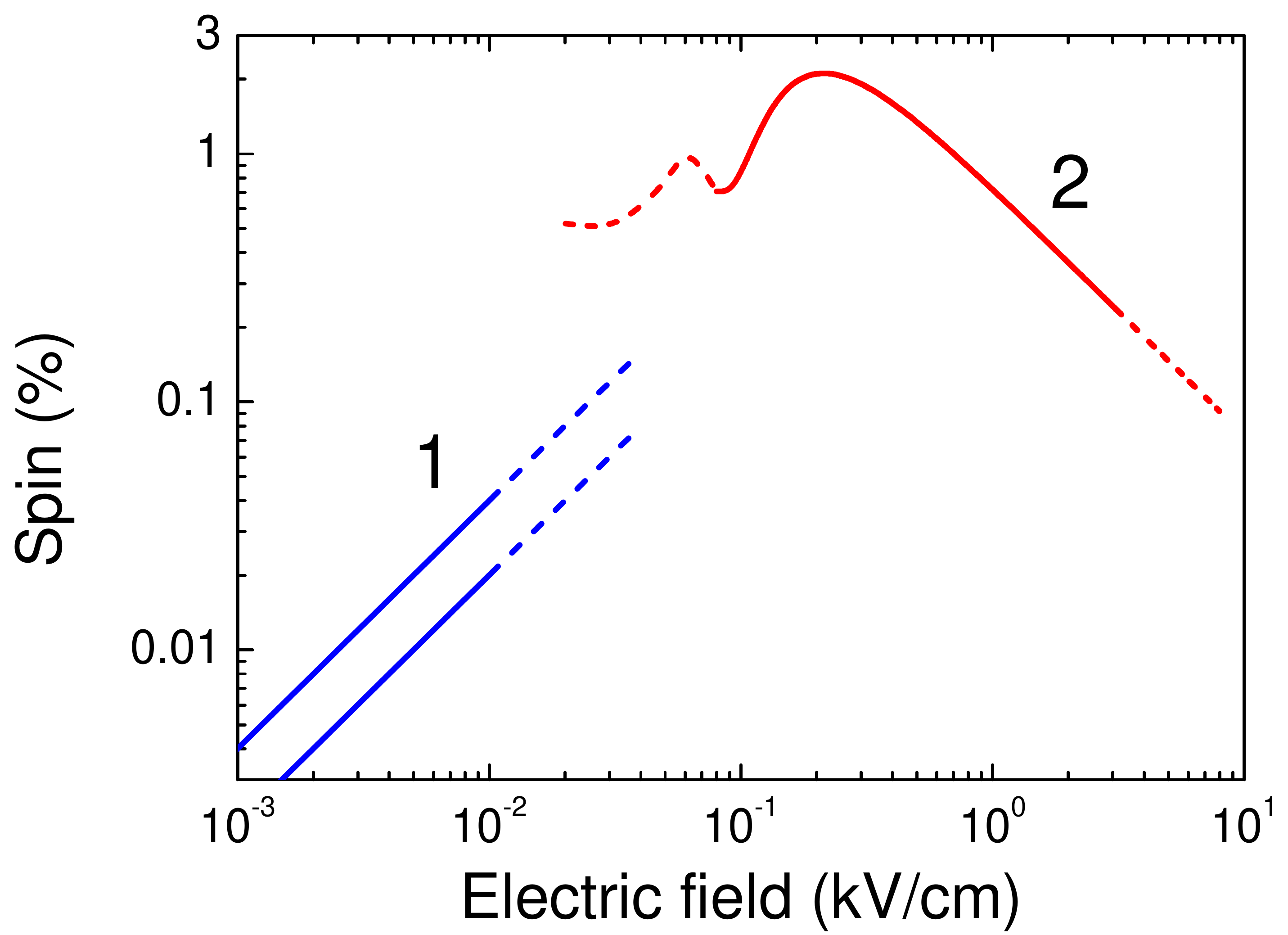}
\caption{Dependence of the current-induced spin on the electric field. Lines~1 are calculated for weak fields in the limits of fast and slow energy relaxation,  curve~2 presents the result of calculation in the streaming regime. Dashed ends of the curve 2 represent the results obtained on the edge of applicability of the streaming model. The parameters used are: $\hbar \beta_{yx}=7$~meV~\AA, $\hbar\omega_0=36$~meV, $m=0.067m_0$, $\tau=0.1$~ps, $\tau_p=10$~ps, $N=10^{11}$~cm$^{-2}$, and zero temperature is assumed.}\label{fig_spin_vs_E}
\end{center}
\end{figure}

In systems with the effective Hamiltonian containing the $\bm p$-linear terms \eqref{H_SO}, the application of an electric field results in orientation of electron spins. Such electrical spin orientation is allowed by the symmetry only in gyrotropic systems where some components of the second-rank pseudotensor $\bm \beta$ are nonzero. In weak fields, this effect has been studied in many works, see the review~\cite{spinorient_PRB}, references therein, and the recent paper~\cite{Shen}. The spin polarization created by a weak electric field is given by
\begin{equation}
\label{s_low_field}
	s_y =  -c \frac{\beta_{yx}}{v} \frac{p_{\rm dr}}{p}\:,
\end{equation}
where $v$ and $p$ are the zero-field root-mean-square electron velocity and momentum, and $p_{\rm dr}=eF\tau_p$. It is remarkable that the coefficient $c$ depends on the ratio of spin and energy relaxation times. 
At slow energy relaxation, the spin is created independently at each energy $\varepsilon$ resulting in a nonequilibrium spin distribution $S_y(\varepsilon)\propto f_0''(\varepsilon)$, where $f_0(\varepsilon)$ is the Fermi-Dirac distribution. In contrast, the fast energy relaxation establishes the equilibrium distribution within each spin-split subband, and the nonequilibrium spin distribution function $S_y(\varepsilon)\propto f_0'(\varepsilon)$. The integration over the energy yields different results for the spin polarization $s_y$. 
For a short-range momentum scattering potential, in two limiting cases of fast and slow energy relaxation the coefficient $c$ in Eq.~\eqref{s_low_field} differs by a factor of two~\cite{spinorient_PRB}: $c_{\rm fast}=1/2$ and $c_{\rm slow}=1/4$. The results for these two extreme cases are shown by straight lines~1 in Fig.~\ref{fig_spin_vs_E}.
One can see that the spin polarization in the fields $F \leq 10$~V/cm does not exceed 0.1\%.

In strong fields where the streaming regime is realized, the situation changes drastically. 
Treating the  term $\beta_{xy}p_y$ in Eq.~\eqref{H_SO} as the spin-orbit splitting of the spectrum we get the shifted needle-like  distributions in the spin subbands with the projection $\pm 1/2$ onto the $x$ axis. Then taking into account the terms $\Omega_y=2\beta_{yx}p_x/\hbar$ and $\Omega_x=2\beta_{xy}p_y/\hbar$ as the components of the precession frequency and neglecting elastic scattering processes in the passive area, we obtain the spin polarization in the form~\cite{NJP}
\begin{equation} \label{synew}
	s_y = - {1 \over 4 \omega_0 \tau} {{\rm Si}(\xi)\over \xi} {\rm sign} (\beta_{yx})\:,
\end{equation}
where $\xi = p_0\sqrt{2|\beta_{yx}|/\pi\hbar e F}$ and ${\rm Si}$ stands for the Fresnel sine integral.
For small values of $\xi$ the ratio ${\rm Si}(\xi) / \xi$ is approximated by $\pi \xi^2/6$ and Eq.~(\ref{synew}) turns into ${s_y = -m\beta_{yx}/ 6 e F \tau}$. 

It follows from Eq.~\eqref{synew} that
the electrically oriented spin is a non-monotonous function of the field strength in the streaming regime. 
The dependence~\eqref{synew} is shown by curve~2 in Fig.~\ref{fig_spin_vs_E}.
The values of the spin orientation have an order of a few percent which is more than an order of magnitude larger than those at low fields. We note that the decrease in the spin relaxation time caused by elastic scattering and described by Eqs.~(\ref{elasttaus}), (\ref{totalsr}) is compensated by a corresponding increase in the generation rate.

\section{Spin beats in the streaming regime}

Consider a undoped 2D structure excited by a short interband pulse creating photoelectrons at the bottom of the lowest conduction subband. Assuming a homogeneous pulse,  the non-equilibrium distribution at the initial moment $t=0$ is given by
\begin{equation} \label{tinit}
f_{\bm p} (t=0) = 2 (\pi \hbar)^2 N \delta(p_x) \delta(p_y)\:,
\end{equation}
where $N$ is the photoelectron density. In the ideal streaming-regime model, the kinetic evolution of the photoelectrons is also described by a product of two delta-functions
\begin{equation} \label{tinitt}
f_{\bm p} (t) = 2 (\pi \hbar)^2 N\ \delta\left(p_x - |eF|t_{\rm tr}\left\{ {t \over t_{\rm tr}}\right\} \right) \delta(p_y)\:,
\end{equation}
with the variable of one function being a function of time with the period $t_{\rm tr}$. Here the symbol $\{ \nu \}$ means the fractional part of the number $\nu$. If the optical pulse is circularly polarized then the generated carriers are spin oriented, and the spin energy distribution is oscillating in time as well. Neglecting the splitting of spin states and the spin relaxation we can supplement Eq.~(\ref{tinitt}) by the similar equation for the time-dependent spin distribution function, cf.~Eq.~\eqref{S_p_f}
\begin{equation} \label{tinits}
\bm S_{\bm p}(t) = {\bm s}_0 (2\pi \hbar)^2 N \delta \left( p_x - |eF|t_{\rm tr}\left\{ {t \over t_{\rm tr}}\right\} \right) \delta(p_y)\:,
\end{equation}
where ${\bm s}_0$ is the initial electron spin per particle. The both functions~(\ref{tinitt}) and~(\ref{tinits}) are periodic until the elastic scattering is ignored. The simplest way to take into account the elastic scattering of the accelerating bunch of electrons is just multiply the above time-dependent functions by the exponential function $\exp{(- t/ \tau_p)}$ turning them into damped oscillatory functions.

The temporary oscillations of $\bm S_{\bm p}(t)$ can be detected in the pump-probe time-resolved experiments based on the spin Kerr or Faraday effects. For the probe light frequency slightly below the quantum-well fundamental edge $E_g/\hbar$, the Kerr rotation angle of polarization plane of the linearly polarized probe pulse is given by
\begin{equation} \label{Kerr}
\Theta_K (t) = \int d \varepsilon \: K(\varepsilon) S_{{\bm p},z}(\varepsilon,t) \:,
\end{equation}
where $\varepsilon$ is the electron energy referred to the bottom of the lowest 2D conduction subband, the energy-dependent coefficient $K(\varepsilon)$ is proportional to $[(E_g - \hbar \omega) (\mu_{eh}/m_e) + \varepsilon]^{-1}$,
$m_e$ is the in-plane electron effective mass, and $\mu_{eh}$ is the reduced mass of an electron and a heavy hole. The substitution of the spin distribution~(\ref{tinits}) multiplied by $\exp{(- t/ \tau_p)}$
into Eq.~(\ref{Kerr}) gives 
\begin{equation} \label{Thetat}
\Theta_K (t) \propto {s_{0,z} \exp{(- t/ \tau_p)} \over (E_g - \hbar\omega)\mu_{eh}/m_e + \hbar\omega_0 \{t/t_{\rm tr} \}^2} + \Delta \Theta_K^{\rm elast.} (t)\:,
\end{equation}
where $\Delta \Theta_K^{\rm elast.} (t)$ is a correction arising due to the elastically scattered carriers and proportional to the ratio $t_{\rm tr}/\tau_p$. While deriving Eq.~(\ref{Thetat}) we took into account that the electron energy in the bunch varies in time as $\varepsilon(t) = p_x^2(t)/2 m = (p_0^2/2m) \{ t/t_{\rm tr}\}^2 \equiv \hbar \omega_0 \{ t/t_{\rm tr}\}^2$.

\begin{figure}[h]
\begin{center}
\includegraphics[width=0.45\linewidth]{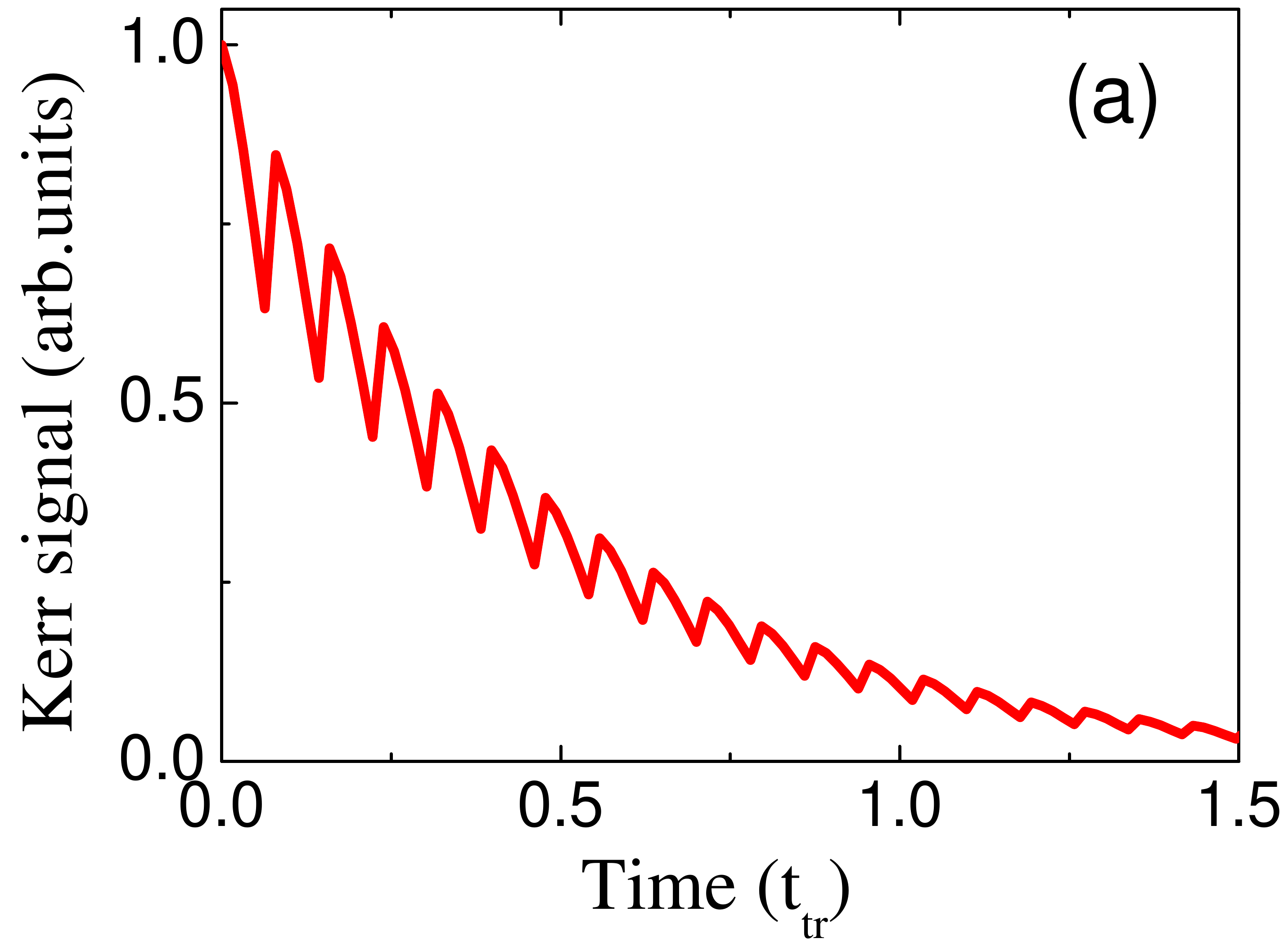}
\quad
\includegraphics[width=0.45\linewidth]{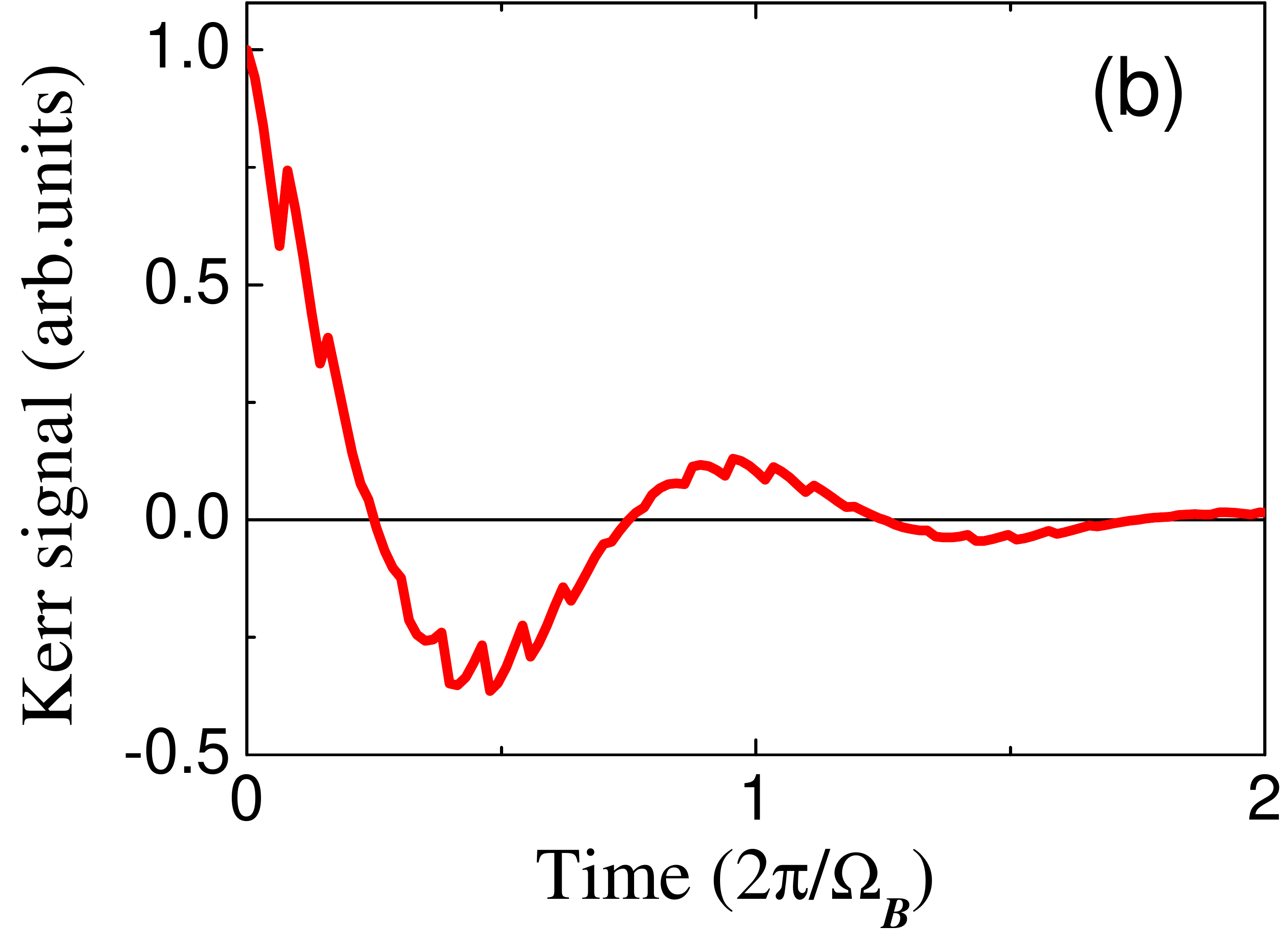}
\caption{The damped oscillatory spin-Kerr signal at ${(E_g - \hbar\omega)\mu_{eh}/m_e =1.7 \hbar\omega_0}$, ${\tau_p=3 \, t_{\rm tr}}$  calculated neglecting the spin-orbit splitting (a) at zero magnetic field, and (b) in the external magnetic field with $\Omega_{\bm B}t_{\rm tr} =0.5$ which corresponds to $B=1.5$~T at $|g|=1$.
}\label{fig_Kerr}
\end{center}
\end{figure}

In the external in-plane magnetic field ${\bm B} \parallel y$ the electron spin component ${\bm s} \perp y$ rotates around the $y$ axis with the angular frequency $\Omega_{\bm B} = g \mu_B B_y/ \hbar$, where $g$ is the electron $g$-factor and $\mu_B$ is the Bohr magneton. The Kerr rotation angle is obtained from Eq.~(\ref{Thetat}) by multiplying the right-hand side by $\cos{(\Omega_{\bm B} t )}$. Indeed, in the absence of the spin-orbit splitting, the spin precession plays a role of the clock independent of the periodic acceleration and emission of optical phonons by the photoelectrons.
The time dependence of the Kerr rotation signal at zero and nonzero magnetic field is plotted in Fig.~\ref{fig_Kerr}, (a) and (b) respectively. Only the first main contribution in Eq.~(\ref{Thetat}) is taken into account in the calculation. The curve in Fig.~\ref{fig_Kerr}(a) has a period $t_{\rm tr}$ whereas the curve in Fig.~\ref{fig_Kerr}(b) has two oscillations with two different periods, the longer one equal to the Larmor precession period $2 \pi/ \Omega_{\bm B}$ and the shorter one equal to the travel time $t_{\rm tr}$.

\section{Spin beats due to the spin-orbit splitting}

\begin{figure}[h]
\begin{center}
\includegraphics[width=0.45\linewidth]{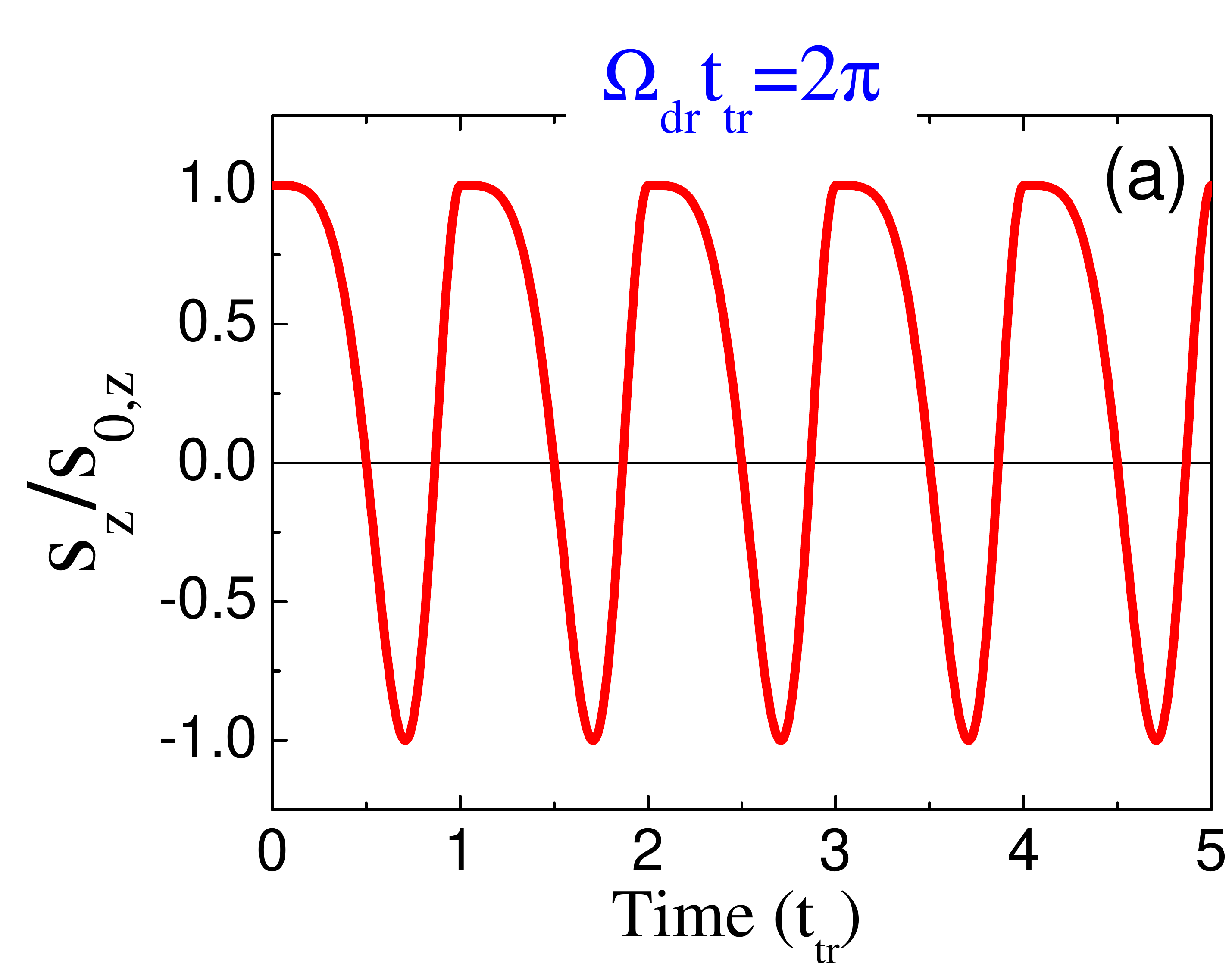}
\quad
\includegraphics[width=0.45\linewidth]{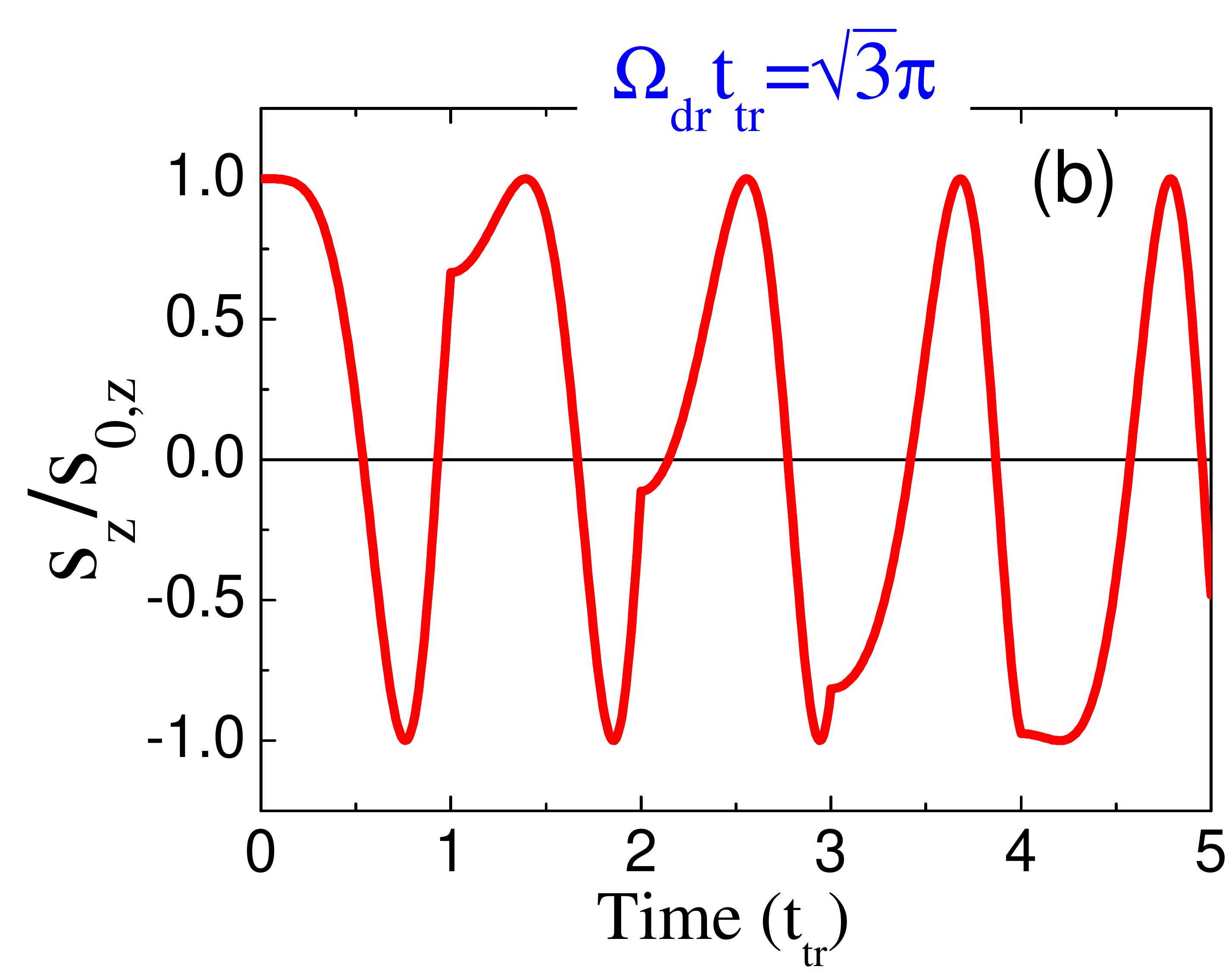}
\\
\includegraphics[width=0.8\linewidth]{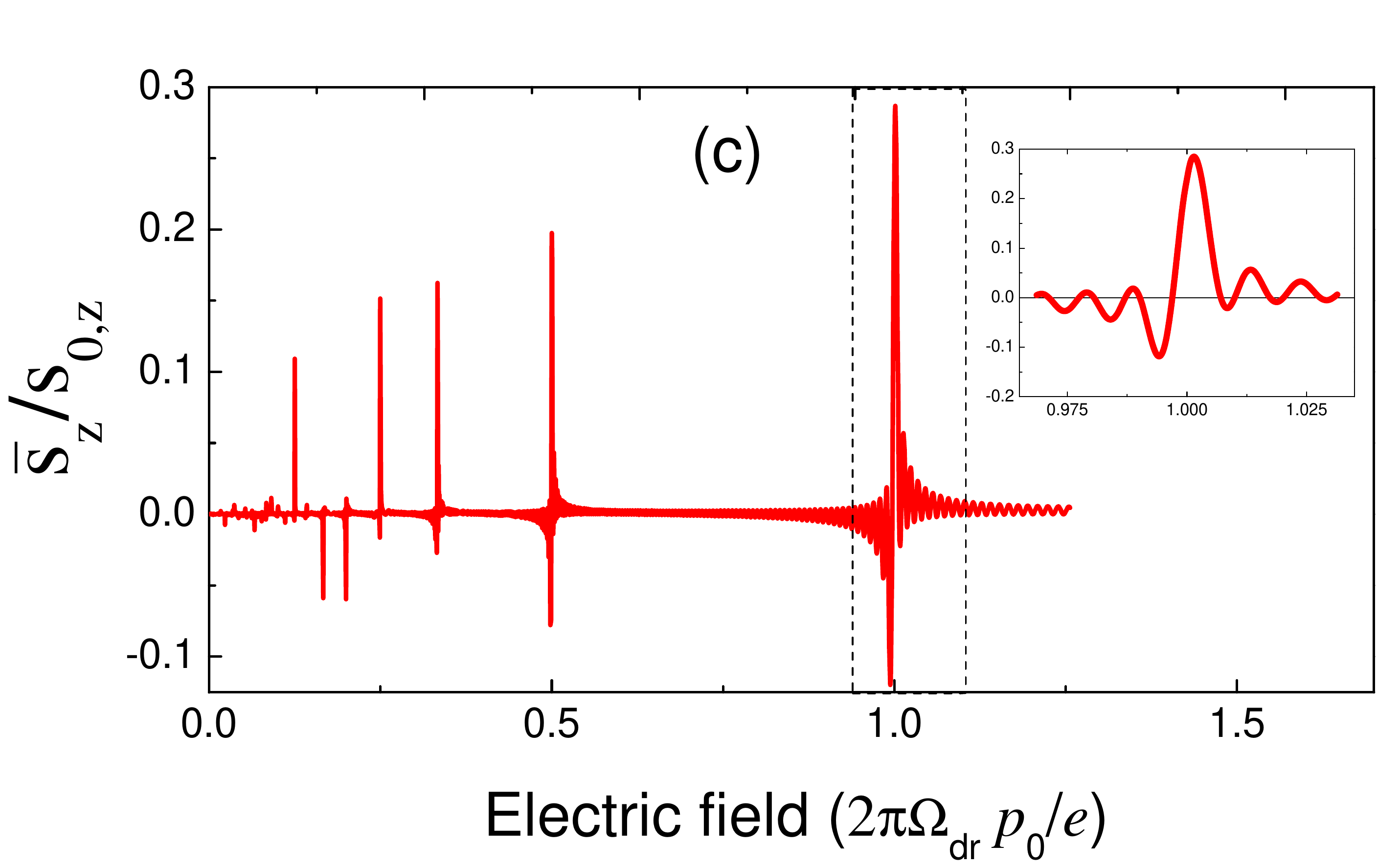}
\caption{(upper panels) Spin beats caused by the spin-orbit splitting for $\Omega_{\rm dr} t_{\rm tr}$ equal to $2 \pi$ (a) and $\sqrt{3} \pi$ (b); (lower panel, c) the time-averaged spin 
$\bar{s}_z(T)$ at $T=100 \, t_{\rm tr}$ calculated in the streaming regime neglecting elastic scattering. Inset highlights the narrow region of electric fields marked by dashed lines.} 
\label{fig_spin_beats_el_field}
\end{center}
\end{figure}

We consider again optical creation of the spin $\bm s \parallel z$ at the bottom of the conduction band at the moment $t=0$. The time-dependent spin distribution is then given by
\begin{equation}
\label{S_p}
\bm S_{\bm p}(t) =N {\bm s}(t) (2\pi \hbar)^2 \delta \left( p_x - |eF|t_{\rm tr}\left\{ {t \over t_{\rm tr}}\right\} \right) \delta(p_y) \:.
\end{equation}
Taking into account the spin-orbit splitting~\eqref{H_SO} we obtain
that the spin dynamics in the absence of external magnetic fields is described by~\cite{NJP}
\[
s_z(t) =s_{0,z} \cos{\Phi(t)}\:, \qquad	s_x(t) = s_{0,z} \sin{\Phi(t)}\:,
\]
where $s_{0,z}$ is the initial spin, and the phase $\Phi$ changes in time according to 
\begin{equation} \label{phit}
	\Phi(t)=  \Omega_{\rm dr} t_{\rm tr}   \biggl( \left[ {t \over t_{\rm tr}}\right] + 
\left\{ {t \over t_{\rm tr}}\right\}^2 \biggr),
\end{equation}
with the symbol $[x]$ denoting the integer part of $x$.
Here ${\Omega_{\rm dr} = 2\sum\limits_{\bm p} \Omega_{\bm p,y} f_{\bm p}/N}$ is the spin precession frequency in the electric field. In weak fields it is linear in $F$~\cite{Kalevich_Korenev} while in the streaming regime one has~\cite{NJP}
\[
\Omega_{\rm dr} = {\beta_{yx} p_0 \over \hbar}.
\]
If $\Omega_{\rm dr} t_{\rm tr}$ is a rational part of $2 \pi$ then the functions $s_z(t), s_x(t)$ are periodic. In particular, for $\Omega_{\rm dr} t_{\rm tr} = 2 \pi n$ with an integer $n$, the period equals to $t_{\rm tr}/n$. For irrational values of $\Omega_{\rm dr} t_{\rm tr}/2 \pi$, the variation $s_{z,x}(t)$ is aperiodic. The time dependence $s_z(t)$ for these two cases is shown in Fig.~\ref{fig_spin_beats_el_field}.

According to Eq.~(\ref{phit}) the phase $\Phi$ is a quadratic function of $\{ t/t_{\rm tr} \}$, this explains why in Fig.~\ref{fig_spin_beats_el_field} the variation of oscillatory function $s_z(t)$ in positive and negative areas is not antisymmetric. For the average spin 
\[
\bar{\bm s}(T) = {1\over T}\int\limits_0^T \bm s(t) dt
\]
we find a sharp electric-field dependence which can be called ``mode-locking''. Introducing $M=[T/t_{\rm tr}]$ and $\varphi = \Omega_{\rm dr} t_{\rm tr}$, we obtain
\begin{equation}
\label{s_av}
	{\bar{s}_z(T) \over s_{0,z}}  = {1\over M} \sum_{k=0}^M \int\limits_0^1 dt \cos{(k\varphi+\varphi t^2)} 
	= {P_1 {\rm Ci}\left(\sqrt{2\varphi/\pi}\right) 
	+ P_2 {\rm Si}\left(\sqrt{2\varphi/\pi}\right) \over 2 M \sin{\varphi}\sqrt{2\varphi/\pi}}, 
\end{equation}
where
\[
P_1 = \sin{M\varphi}+\sin{(M+1)\varphi}+\sin{\varphi}\:,
\qquad
P_2 = \cos{M\varphi}+\cos{(M+1)\varphi}-\cos{\varphi}-1\:.
\]
The expression for $\bar{s}_x(T)$ has a similar form.
One can see that the average spin $\bar{s}_z$ is nonzero if the following condition is fulfilled:
\begin{equation}
	\varphi= \Omega_{\rm dr} t_{\rm tr} = 2\pi n \:,
	\qquad \mbox{i.e.} 
	\qquad F_{n} 
=  {\beta_{yx} p_0^2\over 2\pi \hbar e n }.
\end{equation}
In this case the spin makes $n$ full turns around the $y$ axis during the time $t_{\rm tr}$, and the average spin polarization has a high value. The oscillatory behaviour of $s_z(t)$ in Fig.~\ref{fig_spin_beats_el_field}(a) corresponds to $n=1$. Within the period for $t$ lying between $m t_{\rm tr}$ and $(m+1) t_{\rm tr}$ ($m=0,1,2\dots$) the spin $z$-component is decreasing from its initial value $s_{0,z}$, reaches the minimum value $- s_{0,z}$ and rapidly increases up to $s_{0,z}$. Nonlinear variation of $s_z(t)$ with $\{ t/t_{\rm tr} \}$ results in a nonvanishing value of the average $\bar{s}_z(T)$. Moreover, for $T \gg t_{\rm tr}$, the average
$\bar{s}_z(T)$ is still remarkable in the close proximity of $\varphi$ to $2 \pi n$ where $|\varphi-2\pi n| \ll 1/M$. In the rest area of $\varphi$ values the 
$\bar{s}_z$ have the order of magnitude of $1/M \ll 1$, see Eq.~\eqref{s_av}. The electric field dependence of $\bar{s}_z(T)$ for $T = 100 \: t_{\rm tr}$ is shown in the lower  panel of Fig.~\ref{fig_spin_beats_el_field}, the inset in Figure represents the selected area around $\Omega_{\rm dr} t_{\rm tr} = 2 \pi$ in the enlarged scale.

\begin{figure}[t]
\begin{center}
\includegraphics[width=0.45\linewidth]{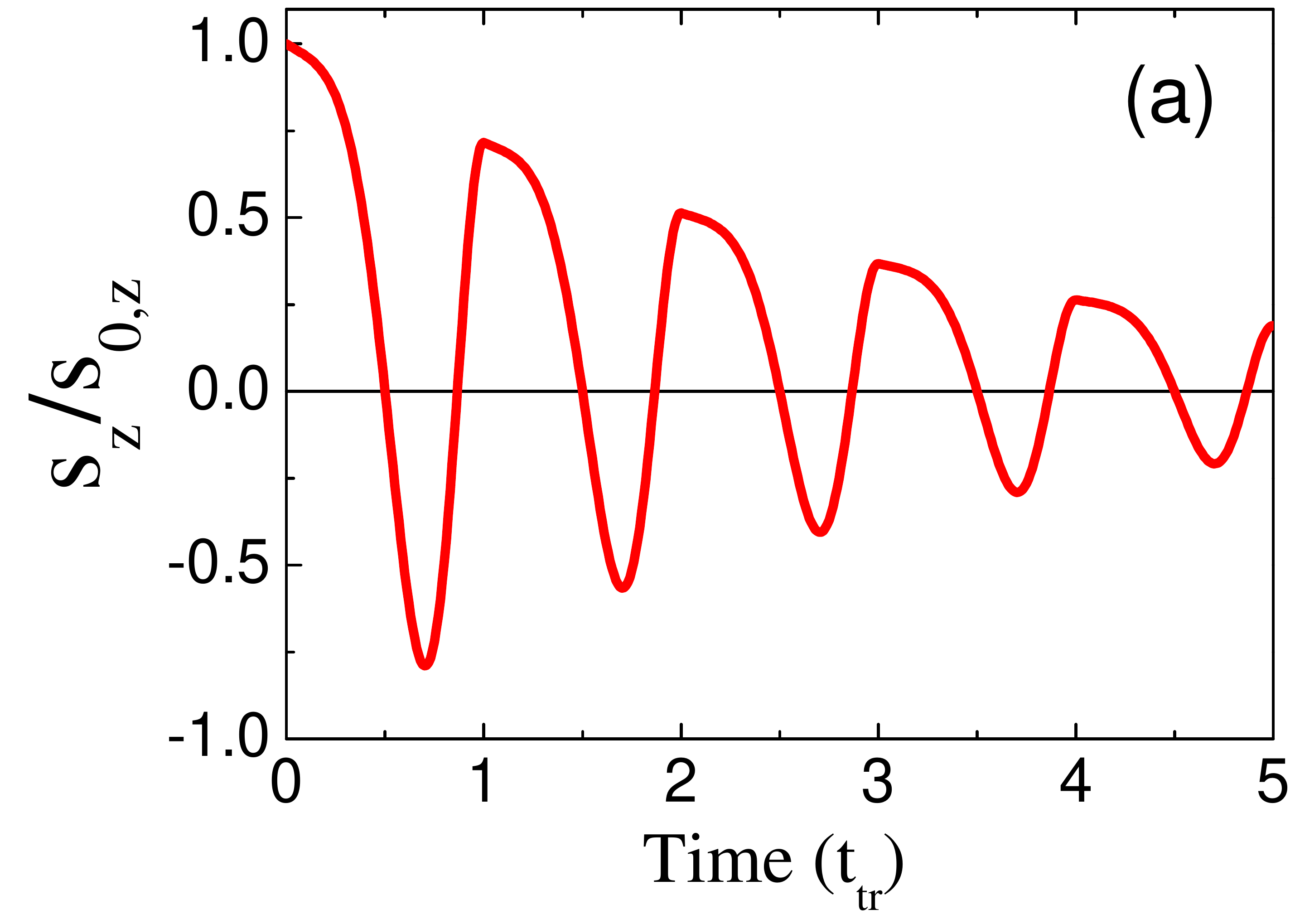}
\quad
\includegraphics[width=0.45\linewidth]{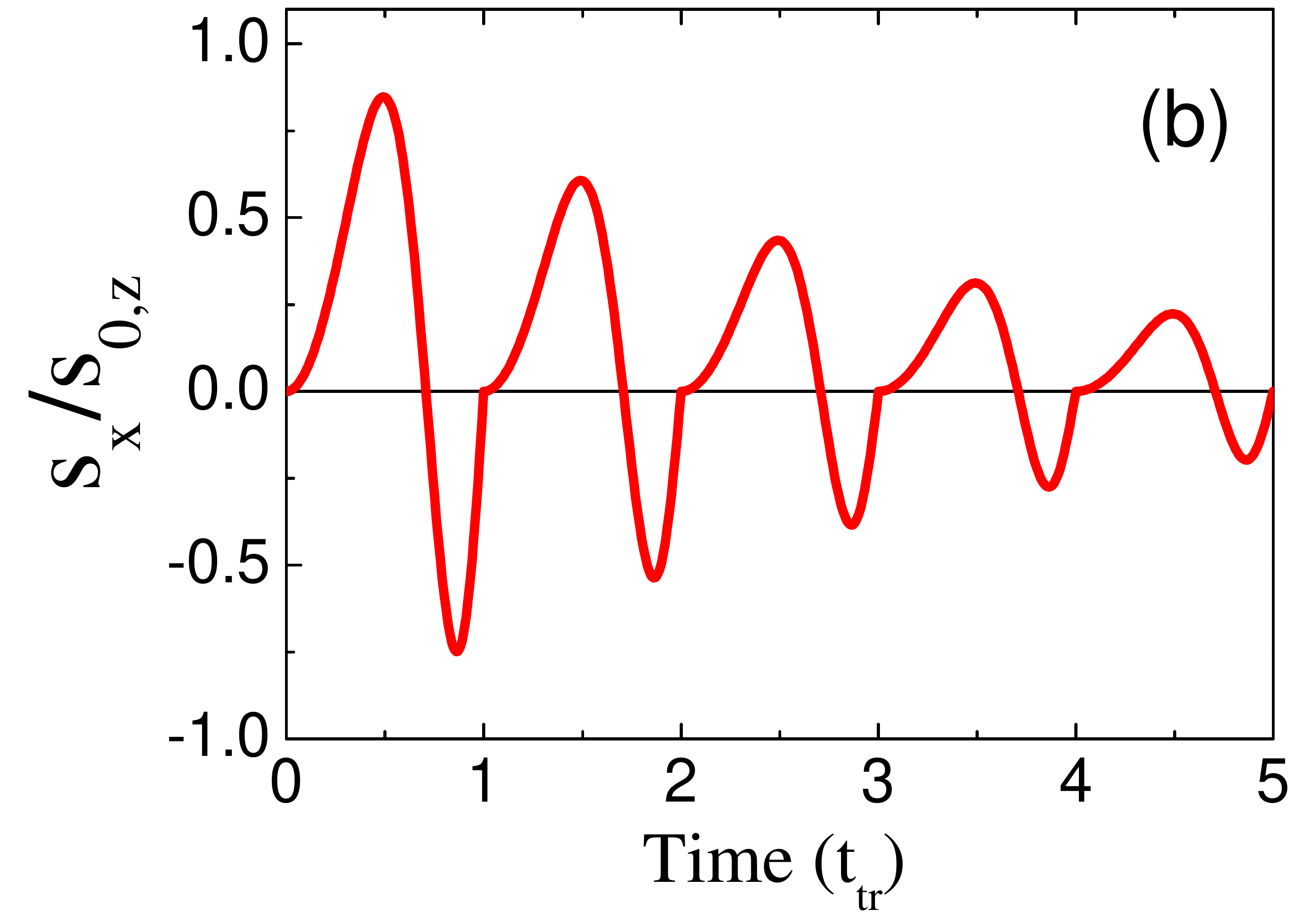}
\caption{Spin beats of $s_z$ (a) and $s_x$ (b) with account for elastic scattering processes at $\Omega_{\rm dr}t_{\rm tr}=2\pi$, $\tau_p/t_{\rm tr}=3$.}
\label{fig_spin_beats_el_field_damp}
\end{center}
\end{figure}

Allowance for other mechanisms of electron scattering and spin relaxation leads to a decay of the oscillation amplitude.
Figure~\ref{fig_spin_beats_el_field_damp} demonstrates the effect of elastic scattering on the dynamics of $s_z$ and $s_x$ components which is taken into account by the factor $\exp{(-t/\tau_p)}$. The Kerr rotation angle is given by Eq.~\eqref{Thetat} where $s_{0,z}$ is replaced by $s_z(t)$.

\section{Conclusion}

In this chapter we have demonstrated the rich variety of spin-dependent phenomena in the streaming regime that can be realized in moderate electric fields in semiconductor heterostructures. The proposed experiments, both steady-state and non-stationary, seem promising for studies of spin properties of two-dimensional electron (or hole) gas.
Observation of photocarrier spin dynamics in the streaming regime by the pump-probe technique would allow the direct access to the time-dependent acceleration of charge carriers in the electric field.

{\bf Acknowledgments}. 
Financial support of RFBR, RF President Grant NSh-1085.2014.2, and 
EU project POLAPHEN is gratefully acknowledged.

\label{lastpage-01}

\end{document}